%% file: main.tex
\documentclass[conference]{IEEEtran}
\IEEEoverridecommandlockouts
\usepackage{booktabs}
\usepackage{epsfig}
\usepackage{latexsym}
\usepackage{multirow}
\usepackage{stfloats}
\usepackage{epstopdf}
\usepackage{tabularx} 
\usepackage{enumerate}
\usepackage{array}
\usepackage{color}
\usepackage{bbm}
\usepackage{caption}
\usepackage{bm}
\usepackage[tight,footnotesize]{subfigure}
\usepackage{balance}

\usepackage{cite}
\usepackage{amsmath,amssymb,amsfonts}
\usepackage{algorithmic}
\usepackage{graphicx}
\usepackage{textcomp}
\usepackage{xcolor}
\usepackage{authblk}
\usepackage{enumitem}
\def\BibTeX{{\rm B\kern-.05em{\sc i\kern-.025em b}\kern-.08em
    T\kern-.1667em\lower.7ex\hbox{E}\kern-.125emX}}

\makeatletter
\newcommand{\linebreakand}{
    \end{@IEEEauthorhalign}
    \hfill\mbox{}\par
    \mbox{}\hfill\begin{@IEEEauthorhalign}
}
\makeatother

\begin{document}

\title{300~GHz Wideband Channel Measurement and Analysis in a Lobby}

\author{Yiqin Wang}
\author{Yuanbo Li}
\author[2]{Yi Chen}
\author[2]{Ziming Yu}
\author{Chong Han}
\affil{Terahertz Wireless Communications (TWC) Laboratory, Shanghai Jiao Tong University, China.\authorcr Email: \{wangyiqin, yuanbo.li, chong.han\}@sjtu.edu.cn}
\affil[2]{Huawei Technologies Co., Ltd., China. Email: \{chenyi171, yuziming\}@huawei.com}


\maketitle

\input{contents/abstract}
\input{contents/introduction}
\input{contents/campaign}
\input{contents/results}
\input{contents/conclusion}


\balance

\bibliographystyle{IEEEtran}
\bibliography{bibliography}

\end{document}

%% file: contents/abstract.tex
\begin{abstract}
The Terahertz (0.1-10~THz) band has been envisioned as one of the promising spectrum bands to support ultra-broadband sixth-generation (6G) and beyond communications. In this paper, a wideband channel measurement campaign in a 500-square-meter indoor lobby at 306-321~GHz is presented. The measurement system consists of a vector network analyzer (VNA)-based channel sounder, and a directional antenna equipped at the receiver to resolve multi-path components (MPCs) in the angular domain. In particular, 21 positions and 3780 channel impulse responses (CIRs) are measured in the lobby, including the line-of-sight (LoS), non-line-of-sight (NLoS) and obstructed-line-of-sight (OLoS) cases. The multi-path characteristics are summarized as follows. First, the main scatterers in the lobby include the glass, the pillar, and the LED screen. Second, best direction and omni-directional path losses are analyzed. Compared with the close-in path loss model, the optimal path loss offset in the alpha-beta path loss model exceeds 86~dB in the LoS case, and accordingly, the exponent decreases to 1.57 and below. Third, more than 10 clusters are observed in OLoS and NLoS cases, compared to 2.17 clusters on average in the LoS case. Fourth, the average power dispersion of MPCs is smaller in both temporal and angular domains in the LoS case, compared with the NLoS and OLoS counterparts. Finally, in contrast to hallway scenarios measured in previous works at the same frequency band, the lobby which is larger in dimension and square in shape, features larger path losses and smaller delay and angular spreads.
\end{abstract}

%% file: contents/introduction.tex
\section{Introduction}

The Terahertz (0.1-10~THz) band has been envisioned as one of the promising spectrum bands for ultra-broadband sixth-generation (6G) and beyond communications~\cite{akyildiz2022terahertz,chen2021terahertz,rappaport2019wireless}. Looking at 2030 and beyond, wireless systems are anticipated to achieve data rates up to terabits per second (Tbps). Compared with the popular millimeter-mave (mmWave) band (30-300~GHz) with the bandwidth of several gigahertz (GHz), the THz band features the broad bandwidth ranging from tens of GHz to several THz, which makes it potential to address the spectrum scarcity and capacity limitations of current wireless systems.

The design of wireless systems in a new spectrum relies on the full knowledge of wireless channels. The study of the wireless channel is composed of understanding the radio propagation, analyzing channel characteristics, and developing channel models and simulators, the foundation of which is the physical channel measurement.
Channel measurement in the THz band is supported by wideband channel sounders. Among different types of wideband channel sounders, the vector network analyzer (VNA)-based channel sounder is popular due to the large bandwidth up to dozens of GHz and its easy access to well-calibrated laboratory instruments~\cite{channel_tutorial}.

Though the exploration of THz wireless channels is of increasing interest, the implementation of THz channel measurements starts from the ``sub-THz'' band, i.e., 100-300~GHz, which overlaps with the mmWave band~\cite{rappaport2019wireless}. For instance, measurement campaigns were conducted around 140~GHz, 190~GHz, and 220~GHz~\cite{chen2021channel,he2021channel, abbasi2022thz, ju2022sub, dupleich2020characterization}.
By contrast, the channel knowledge above 300~GHz frequencies is relatively sparse. Due to the high path loss and, wireless channel measurements above 300~GHz are limited to measuring scenarios like line-of-sight (LoS)~\cite{khalid2019statistical} or short-range environments like desktop, motherboard, and data center~\cite{eckhardt2019measurements, cheng2020thz}.
Despite this, a small proportion of THz channel measurements were implemented at or above 300~GHz in site-scale environments, for instance, an aircraft carbin~\cite{doeker2022channel}, a courtyard~\cite{undi2021angle} and railway or vehicular communication scenarios~\cite{guan2021channel,eckhardt2021channel}.
Besides, our group conducted channel measurements at 306-321~GHz in two kinds of indoor hallway scenarios~\cite{wang2022thz, li2022channel}.

Nevertheless, in 6G wireless communication, THz communication systems will be used in various scenarios. Therefore, it is still a long way to explore the full knowledge of THz wireless channels in all potential scenarios and promising frequency bands. Specially, the World Radiocommunication Conference 2019 (WRC-19) has identified the frequency bands in the range 275-450~GHz for fixed and land mobile service applications, which provides clear guidance for the THz communication industry~\cite{WRC-19-FINAL-ACTS}.
In this spirit, we present a wideband channel measurement campaign conducted in a typical indoor lobby at 306-321~GHz.
Three cases are investigated, including line-of-sight (LoS), non-line-of-sight (NLoS) and obstructed-line-of-sight (OLoS), with 21 measuring positions and 3780 measured channel impulse responses in total. To differentiate, in the OLoS case, the LoS between the transmitter (Tx) and the receiver (Rx) is tangent to the edge of a pillar, while in the NLoS case, it is fully blocked by pillars. In light of the measurement results, effects on multi-path propagation of typical components in the indoor lobby scenario, i.e., pillars, the LED screen and the glass gate, are explored in detail in both temporal and spatial domains. Besides, statistical THz channel characteristics are analyzed in depth, which demonstrates the smallest power dispersion in the LoS case and the most clusters in the OLoS case.
The results are compared with the counterparts in other indoor scenarios.

The remainder of this paper is organized as follows.
In Sec.~\ref{section: campaign}, the wideband channel sounder system and the measurement deployment are introduced. In Sec.~\ref{section: results}, we present the channel measurement results, including the multi-path propagation in light of clustering results, and statistical channel characteristics. Finally, the paper is concluded in Sec.~\ref{section: conclusion}.

%% file: contents/campaign.tex
\section{Channel Measurement Campaign} \label{section: campaign}

In this section, we introduce our VNA-based channel measurement platform, and describe the 306-321~GHz measurement campaign in the lobby on the ground floor of Longbin building, Shanghai Jiao Tong University (SJTU). Data post-processing procedures are also described.

\subsection{Channel Measurement System}


\begin{table}
  \centering
  \caption{Parameters of the measurement.}
    \begin{tabular}{ll}
    \toprule
    Parameter & Value \\
    \midrule
    Frequency band              & 306-321~GHz \\
    Bandwidth                   & 15~GHz \\
    Sweeping interval           & 2.5~MHz \\
    Sweeping points             & 6001 \\
    Time resolution             & 66.7~ps \\
    Space resolution            & 2~cm \\
    Maximum excess delay        & 400~ns \\
    Maximum path length         & 120~m \\
    Tx antenna gain             & 7~dBi \\
    Tx antenna 3-dB beamwidth & 60$^\circ$ \\
    Rx antenna gain             & 25-26~dBi \\
    Rx antenna 3-dB beamwidth & 8$^\circ$ \\
    Tx height          & 2.5~m \\
    Rx height          & 2~m \\
    Rx azimuth rotation range   & $[0^\circ:10^\circ:360^\circ]$ \\
    Rx elevation rotation range & $[-20^\circ:10^\circ:20^\circ]$ \\
    Average noise floor         & -180~dBm \\
    Dynamic range               & 119~dB \\
    \bottomrule
    \end{tabular}
  \label{tab:system_parameter}
\end{table}

Our channel measurement platform supports the frequency ranging from 260~GHz to 400~GHz.
Details of the system are introduced in our previous works~\cite{wang2022thz, li2022channel}.
In this measurement campaign, we investigate the specific frequency band of 306-321~GHz, which covers the bandwidth of 15~GHz.
The frequency sweeping interval is 2.5~MHz.
To compensate for the high path loss, waveguide and WR2.8 horn antenna are installed at Tx and Rx, respectively. The waveguide has the gain around 7~dBi and 3-dB beamwidth about $60^\circ$, while the antenna has the gain around 25-26~dBi and 3-dB beamwidth of $8^\circ$ at 306-321~GHz.

In the mechanical part
, the THz modules are mounted on rotators. The Tx and Rx are lifted by electric lifters to reach the height of 2.5~m and 2~m, respectively. The Tx is static since the beamwidth of the Tx waveguide can cover the measurement deployment. The Rx scans from $0^\circ$ to $360^\circ$ in the azimuth domain and from $-20^\circ$ to $20^\circ$ in the elevation domain. To receive multi-path components (MPCs), the rotation step is $10^\circ$, which is comparable to the 3-dB beamwidth of the Rx antenna. Key parameters of the measurement are summarized in Table~\ref{tab:system_parameter}.

Before data processing, system calibration is carried out to eliminate the effect of THz modules and cables~\cite{he2021channel}. First, the use of frequency multipliers in THz modules ensures that the low-frequency cables can avoid propagating THz signals. Therefore, the system is free of the error caused by frequency mismatch of signals that are carried by the cables.
Despite this, the measured transfer function still contains the response of the THz modules, which is eliminated by directly connecting the Tx and Rx modules and removing the response of calibration from the original measured response.

\subsection{Measurement Deployment}
\begin{figure}
    \centering
    \subfigure[The photo of the lobby.]{
    \includegraphics[width=0.9\linewidth]{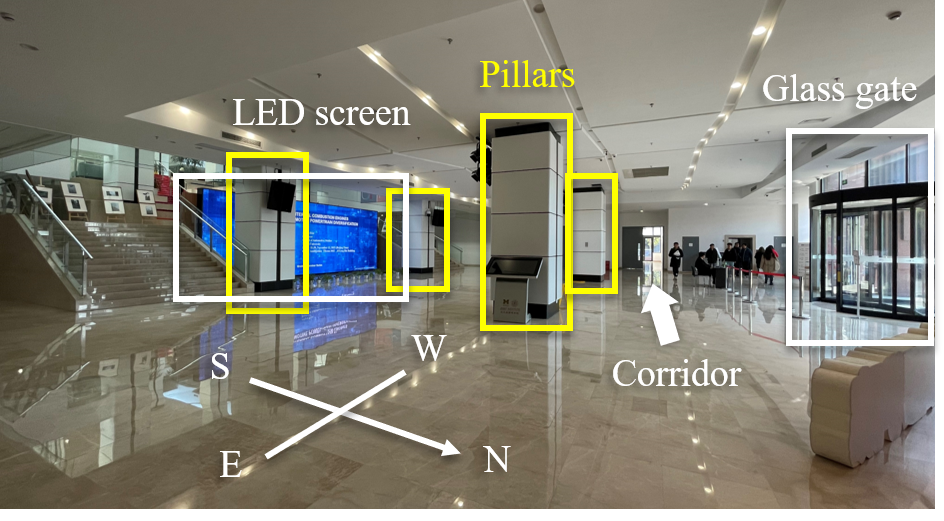}
    }
    \\
    \subfigure[Measurement deployment.]{
    \includegraphics[width=0.95\linewidth]{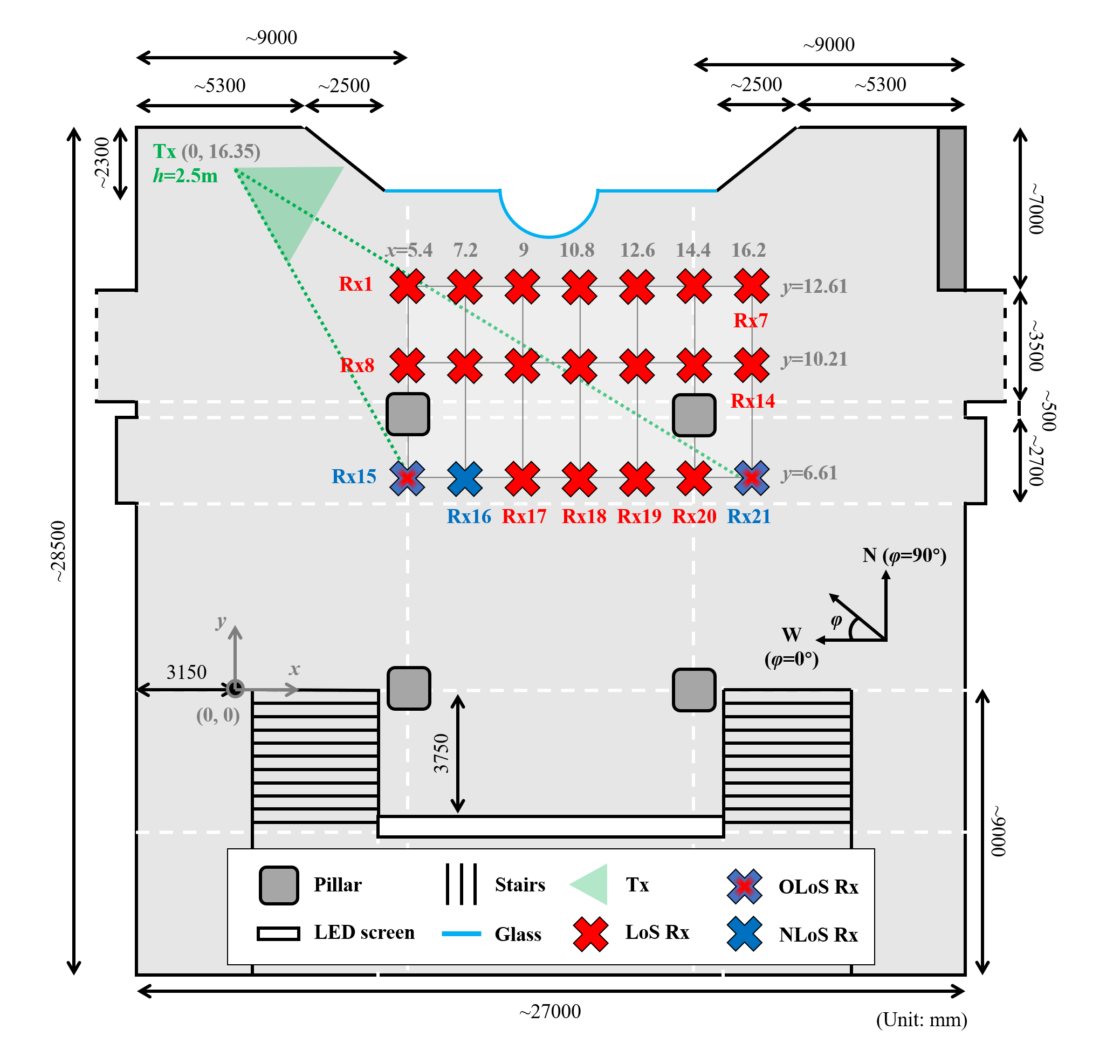}
    }
    \caption{Measurement campaign and deployment in the lobby on the second floor of Longbin building on the SJTU campus.}
    \label{fig:deployment}
\end{figure}

The measurement campaign is conducted in the lobby on the ground floor of the Longbin building on the SJTU campus. As shown in Fig.~\ref{fig:deployment}(a), the major area is the rectangular environment in front of the stairs with more than 500~m$^2$, including a glass revolving gate on the north, four symmetrically distributed pillars in the middle, and one LED screen on the south. Two narrow corridors are extended to the west and east, respectively.
The measurement deployment is illustrated in Fig.~\ref{fig:deployment}(b). Tx is placed at the northwest corner of the lobby, pointing to the center of the 7-by-3 checkerboard-distributed Rx positions. The first two rows of Rx are placed between the glass gate and the pillars. The third row of Rx is placed behind the two pillars. Therefore, 18 LoS Rx positions, 1 NLoS Rx position (Rx16), and 2 OLoS Rx positions (Rx15 and Rx21) are measured in this campaign.

%% file: contents/results.tex
\section{Channel Analysis and Characterization} \label{section: results}
In this section, multi-path propagation analysis in the lobby is presented based on the clustering results.
Moreover, channel characteristics are analyzed in depth, including cluster number, path loss, delay spread and angular spread. Detailed results are summarized in Table~\ref{tab:parameters}, which are derived following the data post-processing procedures as detailed in~\cite{li2022modifiedSAGE}. We clarify that despite the large dynamic range of the measurement system, we focus on statistical characteristics of MPCs within the dynamic range of 30~dB, in order to be compatible with the low dynamic range of general THz devices~\cite{channel_tutorial}. Meanwhile, the noise cutting power thresholds are -160~dB for LoS cases and -165~dB for NLoS cases, i.e., 20~dB higher than the estimated average noise floor from the results of channel impulse response (CIR). The results in the lobby are compared with the counterparts in other indoor scenarios measured in our previous works.



\begin{figure}
    \centering
    \subfigure[MPC distribution at Rx8 (LoS).]{
    \includegraphics[width=0.8\linewidth]{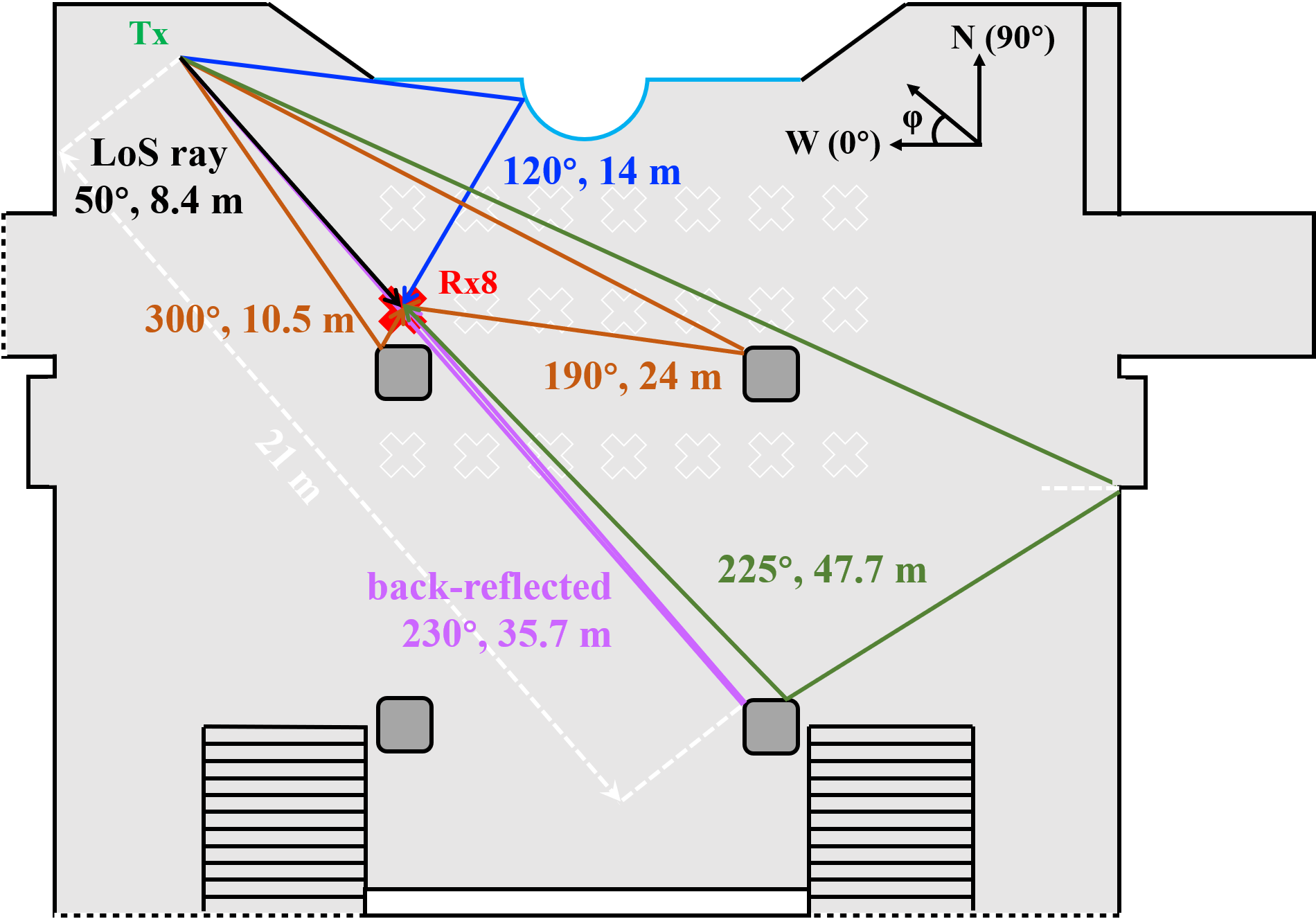}
    }
    \\
    \subfigure[MPC distribution at Rx11 (LoS).]{
    \includegraphics[width=0.8\linewidth]{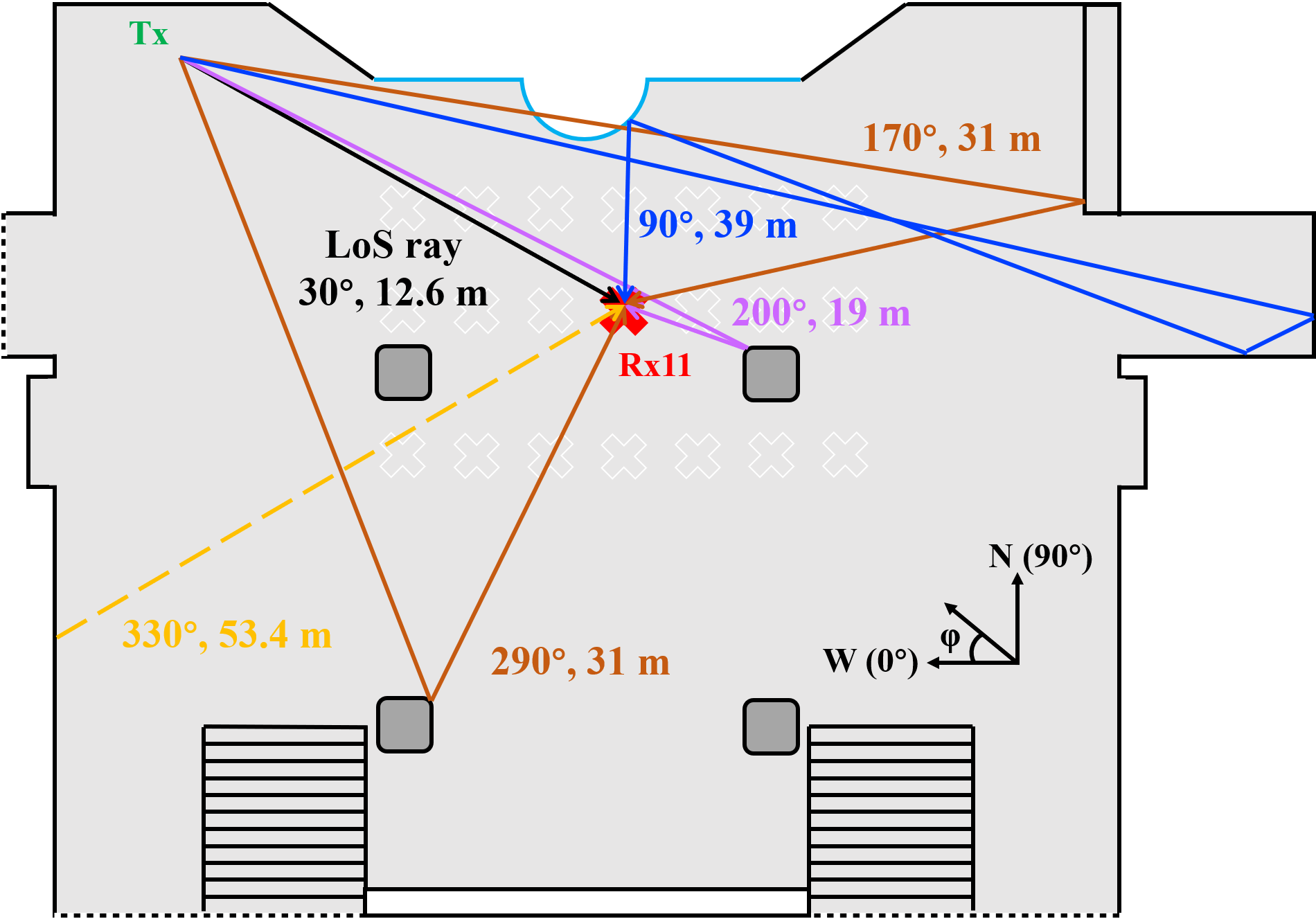}
    }
    \\
    \subfigure[MPC distribution at Rx14 (LoS).]{
    \includegraphics[width=0.8\linewidth]{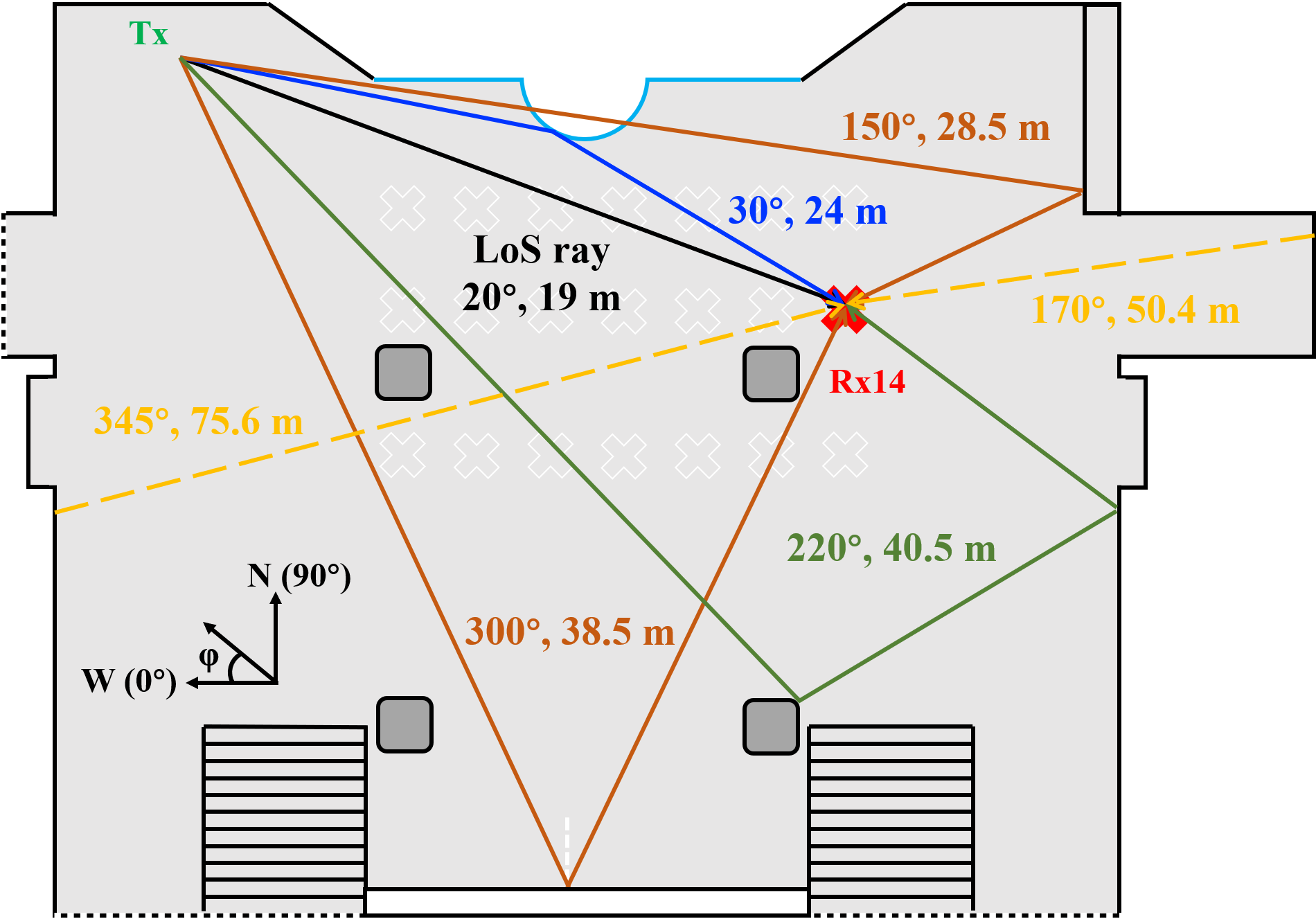}
    }
    \caption{Multi-path propagation analysis at (a) Rx8, (b) Rx11, and (c) Rx14.}
    \label{fig:MPC_analysis_los}
\end{figure}

\subsection{Multi-path Propagation Analysis} \label{sec:MPC_analysis} 
We analyze the multi-path propagation of the THz wave in the indoor lobby scenario based on the clustering results. The distributions of major MPCs at Rx8, Rx11, Rx14, Rx16, and Rx21 are illustrated. Among them, Rx8, Rx11 and Rx14 are chosen as examples with increasing Tx-Rx distances in the LoS case. Rx16 and Rx21 represent NLoS and OLoS positions, respectively.

\begin{figure*}
    \centering
    \subfigure[MPC distribution at Rx16 and Rx21.]{
    \includegraphics[width=0.33\linewidth]{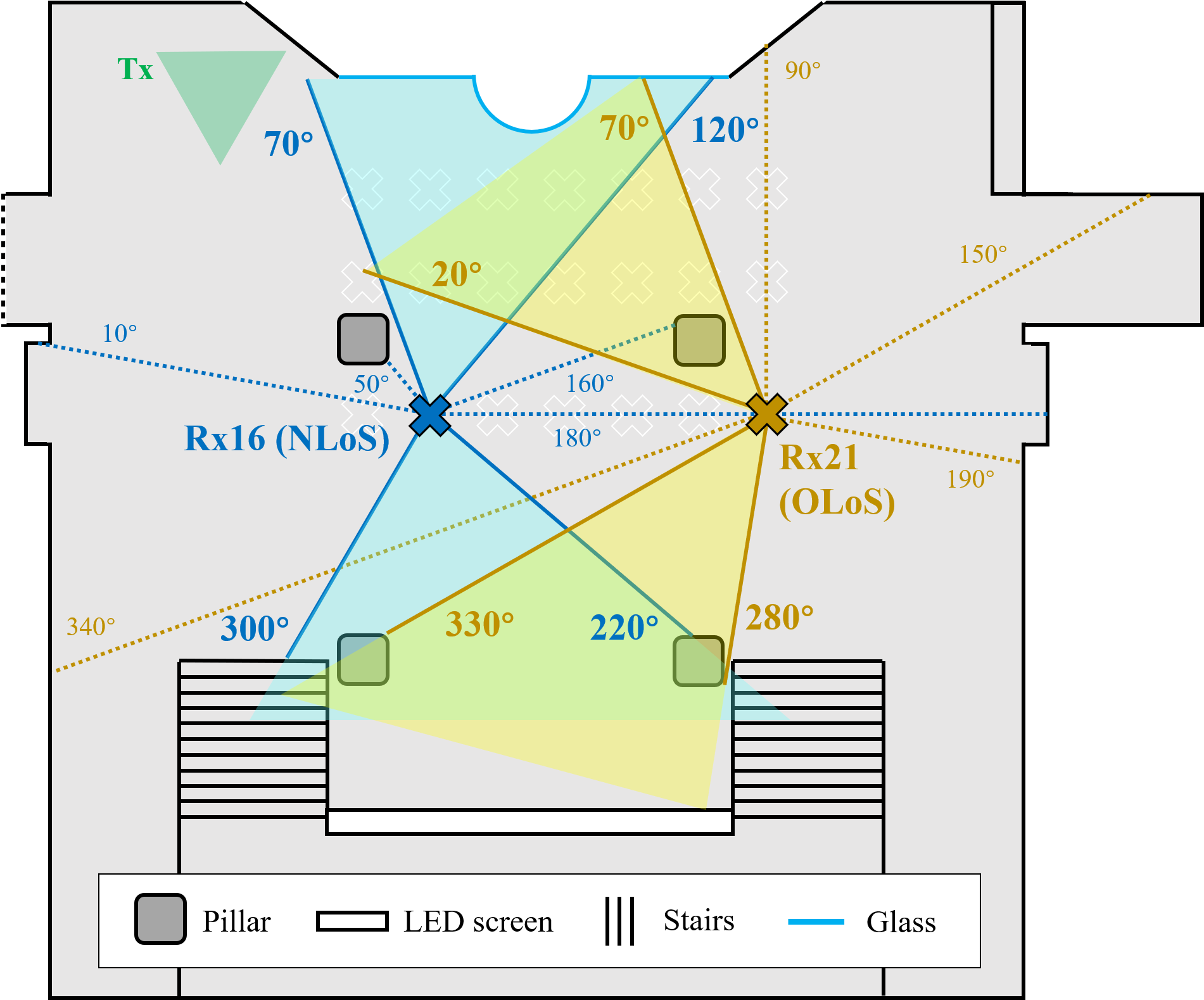}
    }
    \subfigure[Clustering result at Rx16 (NLoS).]{
    \includegraphics[width=0.3\linewidth]{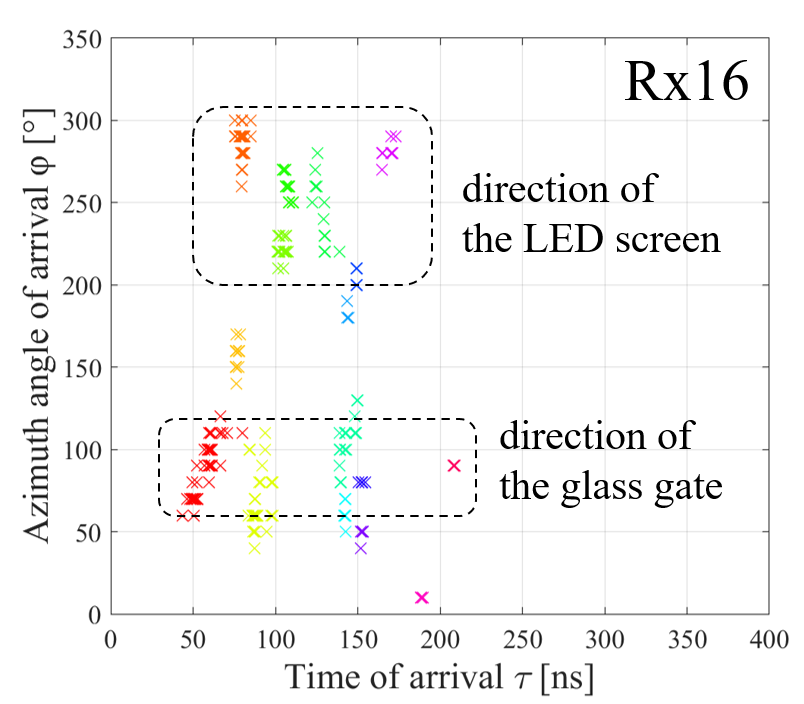}
    }
    \subfigure[Clustering result at Rx21 (OLoS).]{
    \includegraphics[width=0.3\linewidth]{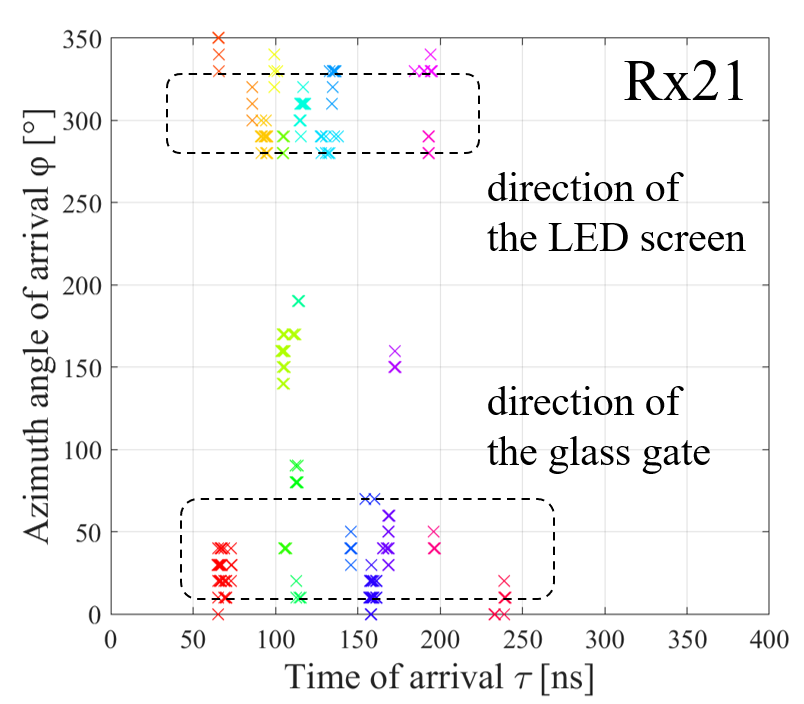}
    }
    \caption{MPC distribution and clustering results at Rx16 (NLoS) and Rx21 (OLoS).}
    \label{fig:MPC_analysis_nlos}
\end{figure*}

In the LoS case as shown in Fig.~\ref{fig:MPC_analysis_los}, the LoS ray, and the specular reflected rays from the pillars, the LED screen and the walls are well observed. The corner of the pillars and the glass gate provide non-specular reflected rays.
For OLoS and NLoS cases, the temporal and spatial distributions of clusters are similar. As shown in Fig.~\ref{fig:MPC_analysis_nlos}, most clusters are received from the directions of the glass gate and the LED screen, with the various delay values ranging from 60~ns to 200~ns. By contrast, very few clusters are received from the west and east walls. To be specific, more than 75\% clusters are received from the glass gate and the LED screen at Rx16 and Rx21, respectively, which demonstrates the abundance of scattering on the glass gate and the LED screen, compared with that on the walls.

\begin{figure}
    \centering
    \includegraphics[width=0.9\linewidth]{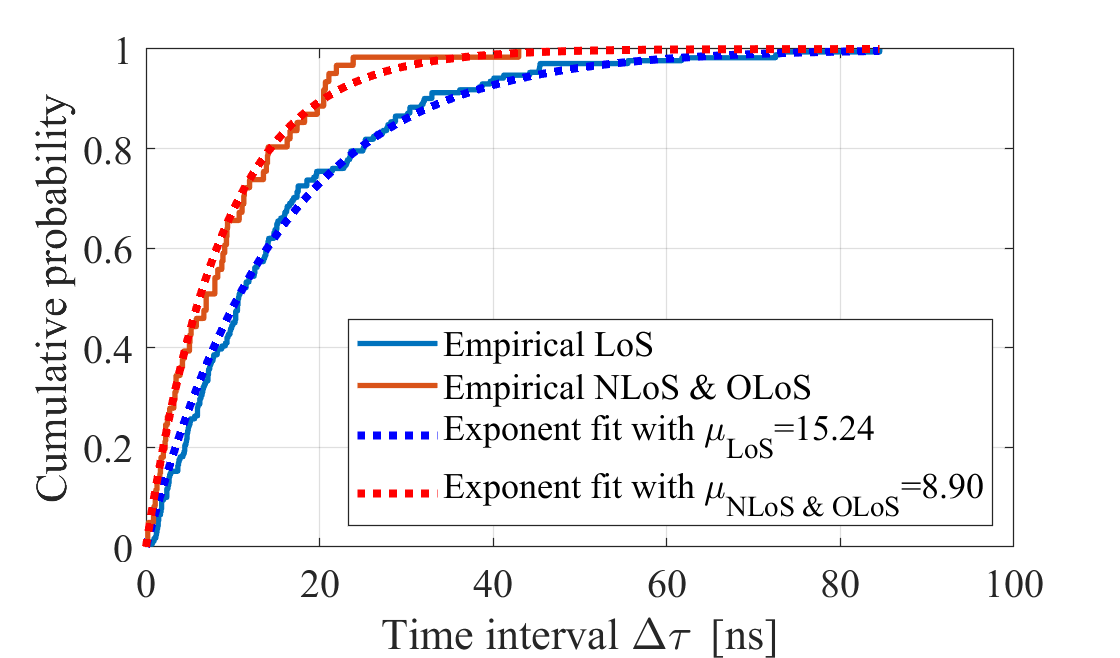}
    \caption{Distribution of the time interval between consecutive clusters.}
    \label{fig:cluster_delay_diff}
\end{figure}

Besides, the temporal arrival of the clusters is modeled by a Poisson process and thus the time interval $\Delta\tau$ between two consecutive clusters is exponentially distributed, as portrayed in Fig.~\ref{fig:cluster_delay_diff}. The average interval between clusters in the LoS case is 15.24~ns. By contrast, the average interval between clusters in OLoS and NLoS cases is 8.9~ns, which demonstrates that OLoS and NLoS receivers capture clusters about twice as frequently as LoS receivers.

\subsection{Cluster Number}
Within the 30~dB dynamic range, the average cluster number is 2 in average in the LoS case, which however increases to 11 and 12 in NLoS and OLoS cases, respectively. In the LoS case, due to the existence of the LoS ray with the highest power, the noise elimination threshold is much higher than those in OLoS and NLoS cases. Therefore, many MPCs are eliminated as noise samples, leading that the number of cluster decreases.
By contrast, in the OLoS case, the cluster that contains the LoS ray and related diffracted paths (named by the ``LoS cluster'') can also be observed, whereas the cluster has a much lower power which is comparable to NLoS clusters. To be specific, the power values of the ``LoS cluster'' received at Rx19 (e.g., far LoS receiver) and Rx15 (the OLoS receiver) are -104~dB and -116~dB, respectively. In comparison, the power of the strongest cluster at Rx16 (the NLoS receiver) is -117~dB. Hence, the OLoS Rx captures not only the attenuated ``LoS cluster'', but also clusters of reflected and scattered rays with weak power, thus containing the most clusters.

\subsection{Path Loss}
Two types of path losses are investigated, namely the best direction path loss and the omni-directional path loss~\cite{wang2022thz}. For each pair of Tx and Rx, the best direction refers to the direction, represented by AoAs, where the MPC has the strongest received power. By contrast, the omni-directional counterpart sums the received power from all directions.

The close-in free space reference distance (CI) and $\alpha$-$\beta$ path loss models~\cite{rappaport2015wideband} are invoked, which are expressed as
\begin{subequations}
\begin{align}
\text{PL}^{\rm CI} &= 10\times\text{PLE}\times\log_{10}\frac{d}{d_0}+\text{FSPL}(d_0)+X^{\rm CI}_{\sigma_{\rm SF}},\\
\text{PL}^{\alpha\beta} &= 10\times\alpha\times\log_{10}d+\beta+X^{\alpha\beta}_{\sigma_{\rm SF}},
\end{align}
\end{subequations}
where $d$ denotes the distance between Tx and Rx. $d_0$, which is 1~m in this work, represents the reference distance. The free-space path loss (FSPL) is given by the Friis’ law. PLE is the path loss exponent, while $\alpha$ is the slope coefficient and $\beta$ represents the optimized path loss offset in dB. $X^{\rm CI}_{\sigma_{\rm SF}}$ and $X^{\alpha\beta}_{\sigma_{\rm SF}}$ are two zero-mean Gaussian random variables with respective standard deviation $\sigma^{\rm CI}_{\rm SF}$ and $\sigma^{\alpha\beta}_{\rm SF}$ in dB, indicating the fluctuation caused by shadow fading. These parameters are determined by model fitting on measured path losses and distances.

\begin{figure}
    \centering
    \subfigure[Path loss measurement and fitted CI models.]{
    \includegraphics[width=\linewidth]{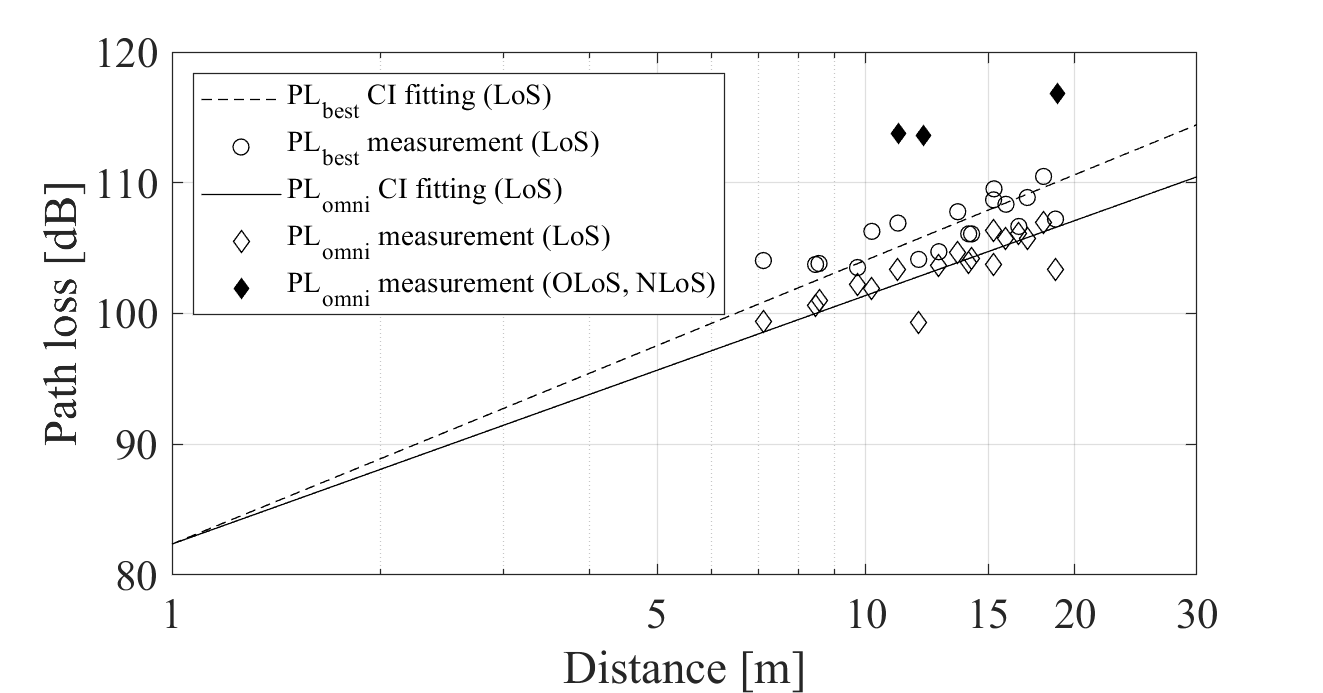}
    }
    \\
    \subfigure[Path loss measurement and fitted $\alpha$-$\beta$ models.]{
    \includegraphics[width=\linewidth]{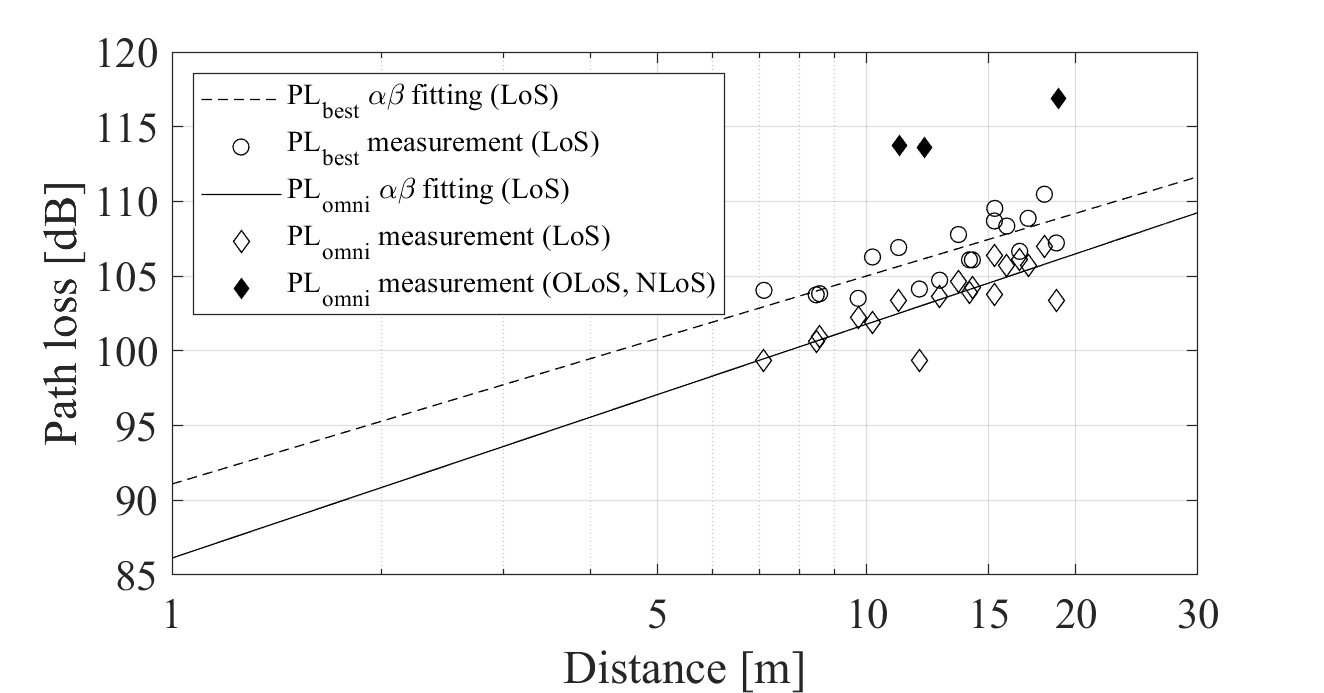}
    }
    \caption{Path loss measurement and fitted (a) CI models, (b) $\alpha$-$\beta$ models.}
    \label{fig:pl}
\end{figure}

As shown in Fig.~\ref{fig:pl}(a), the measurement results yield the LoS PLE valus of 2.17 for the best direction path loss and 1.90 for the omni-direction path loss, respectively.
Observations and justifications are elaborated as follows. First, the best direction PLE is higher than 2, i.e., the PLE for ideal free-space path loss. On one hand, though the beamwidth of Tx can nearly cover the deployment, Rx positions that are away from the center of the transmitted beam experience larger loss than expected. On the other hand, while Tx is static, Rx scans the angular domains with the step of $10^\circ$, which causes possible directional misalignment between Tx and Rx. In other words, when the Rx is aligned to the Tx, the aligned angles are calculated by the coordinates of the transceivers, which are not necessarily the multiple of $10^\circ$ caused by misalignment.
Based on the measurement results, the extra loss can reach 7~dB at most at Rx6, where the deviation in azimuth and elevation AoA reaches $5^\circ$ and $2^\circ$, respectively. The value accords with the radiation pattern of the antenna.

Second, the omni-directional power sums the received power, resulting in a PLE smaller than the best direction PLE. Despite, attributing to the glass gate that leaks power, the path loss is still larger than that specified by 3GPP TR~38.901 for the indoor hotspot (InH)-office LoS scenario~\cite{3gpp38901} as well as those in indoor hallway scenarios based on our measurements in the same frequency band~\cite{wang2022thz,li2022channel}. 

The results of the $\alpha$-$\beta$ model are shown in Fig.~\ref{fig:pl}(b). First, the fact that the optimal path loss offset parameter, $\beta$, takes values that are larger than the reference offset of 82~dB in CI models, which verifies the existence of the extra loss of power. Second, the omni-directional path loss is typically smaller than the best direction path loss by 4-5~dB in the LoS case in the lobby.
Compared with the CI model, the flexibility of the $\alpha$-$\beta$ model is increased due to the extra offset parameter. Nevertheless, the value of the slope parameter is also affected during the optimization. In our case, Tx-Rx distances vary from 6.6~m to 18.9~m, which are concentrated. As a result, a small change of the offset renders a large deviation of the slope. Therefore, further conclusions on model parameters still need to be supported by more measurement data.

\begin{figure}
    \centering
    \includegraphics[width=0.8\linewidth]{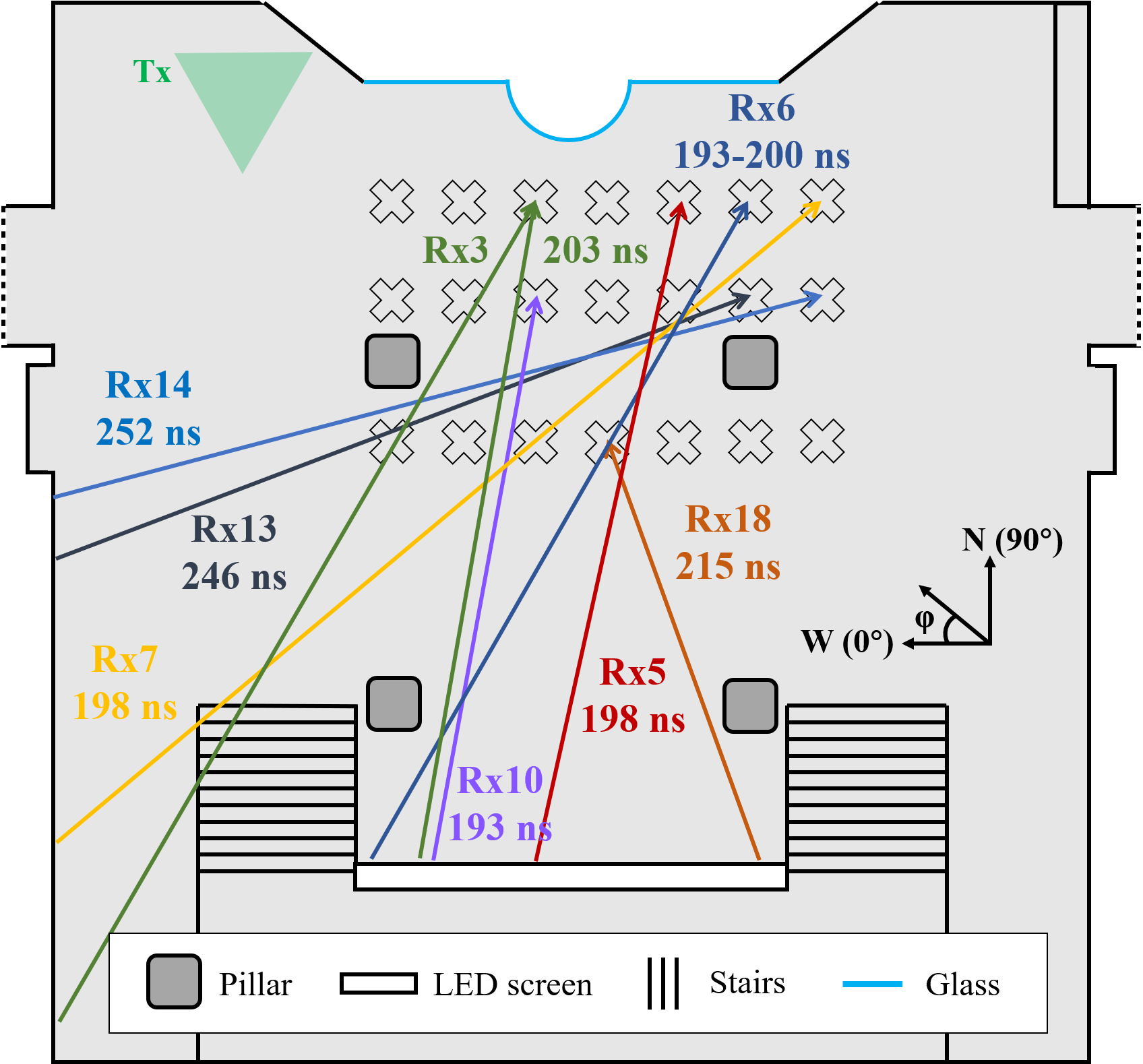}
    \caption{Distribution of long paths with delay larger than 190~ns.}
    \label{fig:long_paths}
\end{figure}

\subsection{Delay and Angular Spreads}
We use the root-mean-square (RMS) delay spread (DS), azimuth spread of angle (ASA) and elevation spread of angle (ESA) to measure the power dispersion of MPCs in temporal and spatial domains, respectively.
First, the values of DS and ASA are smaller in the LoS case, compared with those in OLoS and NLoS cases. This is mainly attributed to the dominance of the LoS ray, which is partially weakened or entirely eliminated in OLoS and NLoS cases. Therefore, the power dominated by reflected and scattered paths is more spreaded in OLoS and NLoS cases, both in temporal and angular domains.
Second, in the LoS case, the delay spread is large at positions where paths with the delay larger than 190~ns are observed. The distribution of these long paths is illustrated in Fig.~\ref{fig:long_paths}. We discover that the detectable long paths typically reach the Rx from the southwest, experiencing twice reflection by the wall or the LED screen, with the azimuth AoA ranging from $280^\circ$ to $350^\circ$.


\begin{table}
\caption{Summary of channel characteristics in the lobby at 306-321~GHz (PL: path loss, DS: delay spread, ASA/ESA: azimuth/elevation spread of angle, CDS: intra-cluster delay spread, CASA/CESA: intra-cluster azimuth/elevation spread of angle).}
\centering
\includegraphics[width=0.95\linewidth]{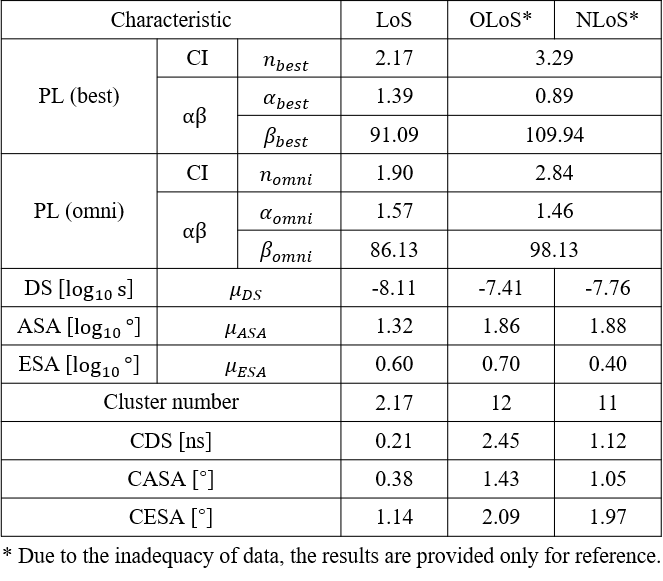}
\label{tab:parameters}
\end{table}

\subsection{Comparison of Characteristics in Indoor Scenarios}
In~\cite{wang2022thz,li2022channel}, we measured wireless THz channels at the same frequency band in indoor hallway scenarios, namely, the L-shaped hallway and the corridor.
In this part, we compare channel characteristics in the lobby and these hallway scenarios. We focus on characteristics in the LoS case due to the inadequacy of NLoS data in the lobby.
First, within the dynamic range of 30~dB as explained in Section~\ref{section: results}, the average numbers of clusters in the LoS case in indoor scenarios is 2-5 at 306-321~GHz, which is much smaller than the counterpart in the InH - Indoor office scenario specified in 3GPP TR~38.901 under 100~GHz.
Second, in the LoS case, the waveguide effect can be observed in hallway scenarios, and therefore, the path losses in the lobby are larger than the counterparts in hallway scenarios.
Third, delay and angular spreads in the LoS case are smaller in the lobby for the two following reasons. One reason is that the dimension of the lobby (19.5$\times$27~m$^2$) is larger than hallways (2.97$\times$32.53~m$^2$ and 1.85$\times$63.88~m$^2$), thus the LoS ray is more dominant in the lobby as reflected and scattered rays travel longer distances and consume the power. Another reason is that long and narrow indoor scenarios like hallways, compared with bordered scenarios whose geometry is analogous to square like the lobby, have strong back-reflection with long delay and opposite angle of arrival against the LoS ray.

%% file: contents/conclusion.tex
\section{Conclusion} \label{section: conclusion}
In this paper, we conducted a wideband channel measurement campaign in an indoor lobby at 306-321~GHz, by using a frequency-domain VNA-based channel sounder.
The indoor scenario is featured by the glass gate, the LED screen, and symmetric pillars in the middle. Based on the analysis of multi-path components and statistical characteristics in the lobby at 306-321~GHz, the following conclusions are drawn.
\begin{itemize}[leftmargin=*]
    \item Specular reflected rays from the pillars, the LED screen and walls, and transmission rays through the glass gate are well observed.
    \item The corner of the pillars and the semi-circular glass gate provide back-reflected and non-specular reflected rays.
    \item The abundance of scattering of the glass gate and the LED screen is demonstrated, which occupies at least 75\% MPC sources in OLoS and NLoS cases.
    \item More than 10 clusters are observed in OLoS and NLoS cases, which are more abundant compared with 2.17 clusters on average in the LoS case.
    \item The CI path loss model indicates that LoS PLEs are close to that for ideal free-space path loss, while the $\alpha$-$\beta$ model yields optimal offsets larger than the reference offset in CI models.
    \item The path loss, average DS and ASA in the LoS case are all smaller than OLoS and NLoS counterparts.
    \item In contrast to hallways, the lobby is larger in dimension and more square in shape, and features larger path losses and smaller delay and angular spreads.
    \end{itemize}
